\documentclass[sigconf, nonacm]{acmart}
\usepackage{multirow}
\AtBeginDocument{%
  }

\usepackage{subcaption}
\setcopyright{acmlicensed}
\copyrightyear{2018}
\acmYear{2018}
\acmDOI{XXXXXXX.XXXXXXX}
\acmConference[Conference acronym 'XX]{Make sure to enter the correct
  conference title from your rights confirmation email}{June 03--05,
  2018}{Woodstock, NY}
\acmISBN{978-1-4503-XXXX-X/2018/06}

\begin{document}

\title{AgentGR: Semantic-aware Agentic Group Decision-Making Simulator for Group Recommendation}


\author{Yangtao Zhou}
\email{zhouyangtao@xidian.edu.cn}
\affiliation{
  \institution{Xidian University}
  \city{Xi'an}
  \country{ China}
}

\author{Wenhao You}
\email{ywhss@stu.xidian.edu.cn}
\affiliation{
  \institution{Xidian University}
  \city{Xi'an}
  \country{ China}
}

\author{Hua Chu}
\email{hchu@mail.xidian.edu.cn}
\affiliation{
  \institution{Xidian University}
  \city{Xi'an}
  \country{ China}
}

\author{Shihao Guo}
\email{shihaoguo@stu.xidian.edu.cn}
\affiliation{
  \institution{Xidian University}
  \city{Xi'an}
  \country{ China}
}

\author{Jianan Li}
\email{lijianan@xidian.edu.cn}
\affiliation{
  \institution{Xidian University}
  \city{Xi'an}
  \country{ China}
}

\author{Zhifu Zhao}
\email{zfzhao@xidian.edu.cn}
\affiliation{
  \institution{Xidian University}
  \city{Xi'an}
  \country{ China}
}

\author{Qingshan Li}
\email{qshli@mail.xidian.edu.cn}
\affiliation{
  \institution{Xidian University}
  \city{Xi'an}
  \country{ China}
}
\authornote{Corresponding author.}

\renewcommand{\shortauthors}{Trovato et al.}

\begin{abstract}
 Group Recommendation (GR) aims to suggest items to a group of users, which has become a critical component of modern social platforms. Existing GR methods focus on aggregating individual user preferences with advanced neural networks to infer group preferences. Despite effectiveness, they essentially treat group preference learning as a simple preference aggregation process, failing to capture the complex dynamics of real-world group decision-making. To address these limitations, we propose AgentGR, a novel Semantic-aware \textbf{\underline{Agent}}ic Group Decision-Making Simulator for \textbf{\underline{G}}roup \textbf{\underline{R}}ecommendations, inspired by the semantic reasoning and human behavior simulation capabilities of LLM-driven agents. It aims to jointly capture collaborative-semantic user preferences for member-role-playing and simulate dynamic group interactions to reflect real-world group decision-making processes, thereby boosting recommendation performance. Specifically, to capture collaborative-semantic user preferences, we introduce a semantic meta-path guided chain-of-preference reasoning mechanism that integrates high-order collaborative filtering signals and textual semantics to improve user preference profiles. To model the complex dynamics of group decision-making, we first recognize group topic and leadership to explicitly model the influencing factors within the group decision processes. Building on these, we simulate group-level decision dynamics via two multi-agent simulation strategies for recommendations: a static workflow-based strategy for efficiency and a dynamic dialogue-based strategy for precision. Extensive experiments on two real-world datasets show that AgentGR significantly outperforms state-of-the-art baselines in both recommendation accuracy and group decision simulation, highlighting its potential for real-world GR applications.
\end{abstract}




\received{20 February 2007}
\received[revised]{12 March 2009}
\received[accepted]{5 June 2009}

\maketitle

\section{Introduction}
Group activities, such as family trips, social gatherings, and club events, are integral to people's daily life \cite{zhou2025spatiotemporal}. As the demand for group-level experiences grows, generating personalized and effective recommendations for a group of users becomes an important task in recommendation systems. Therefore, Group Recommendation (GR) has emerged. GR aims to capture and aggregate diverse individual user preferences to generate recommendations that suit the entire group \cite{Ye_2025,kim2025leveraging}, having garnered significant research attention over the past decade.

\begin{figure}[h]
\centering
\includegraphics[width=\columnwidth]{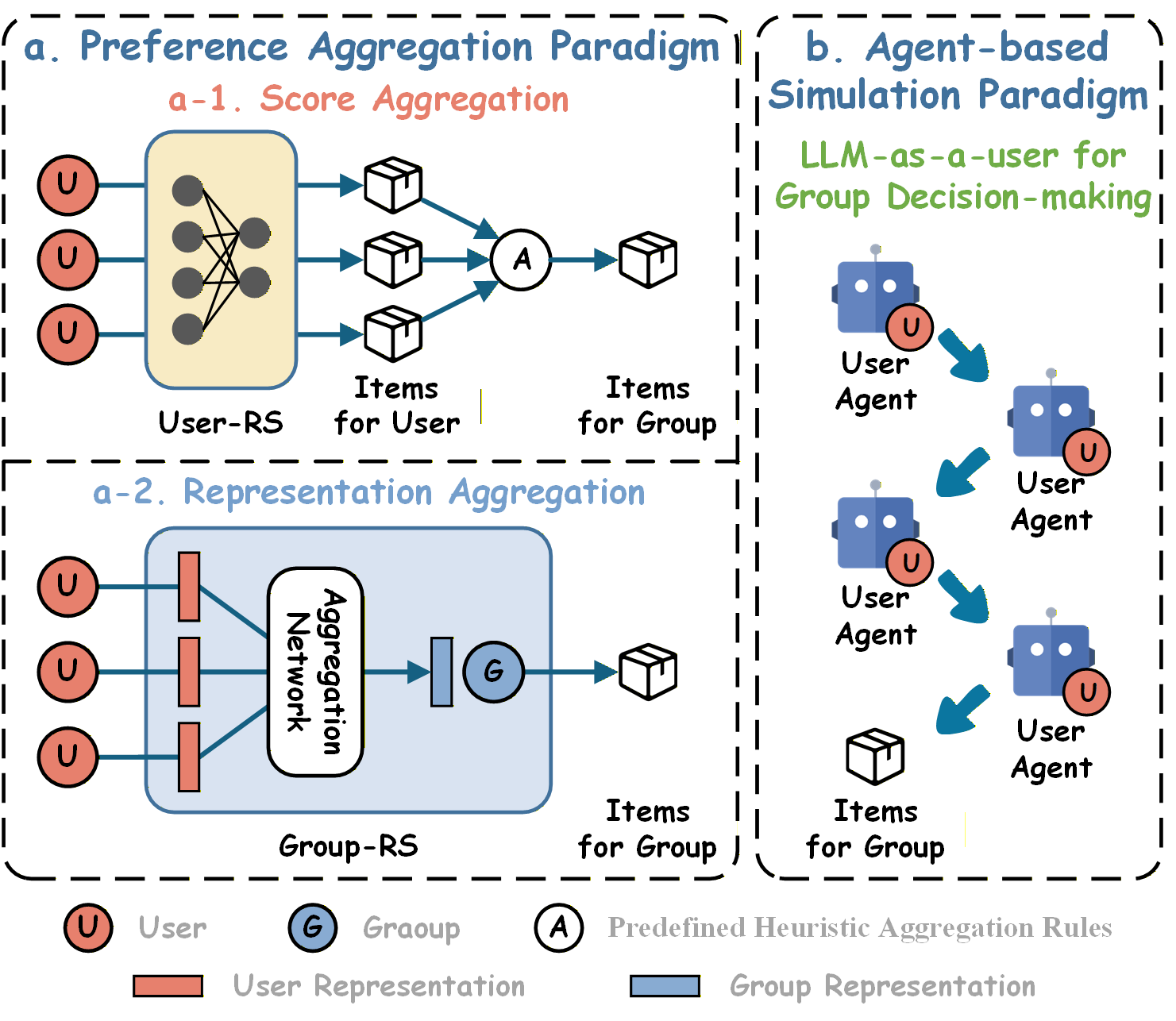} 
\caption{Comparison between the conventional preference aggregation paradigm and our proposed agent-based simulation paradigm.}
\label{fig0}
\end{figure}

Recent studies \cite{10.1145/3209978.3209998,wu2023consrec} have explored advanced neural network architectures to simulate group decision-making processes for group preference learning, achieving promising progress. As shown in Figure~\ref{fig0}.a, existing GR methods can be broadly classified into two categories: score aggregation and representation aggregation. The score aggregation methods use predefined heuristic rules, such as most pleasure or least misery, to aggregate individual recommendation scores into group-level recommendations \cite{baltrunas2010group,boratto2011state}. However, these methods fall short in capturing the complex interactions during group decision-making processes due to their oversimplification and inflexibility \cite{zhou2025spatiotemporal}. To address this, representation aggregation methods have emerged, utilizing advanced deep neural networks for aggregating user preference representations rather than recommendation scores. For example, early attention-based methods introduce attention mechanisms to explicitly model the importance contributions of group members in group preference learning \cite{10.1145/3209978.3209998,deng2021knowledge}. More recently, graph-based methods, particularly hypergraph neural networks, capture high-order relation patterns by aggregating information from the graph structures among groups, users, and items, achieving state-of-the-art performance \cite{jia2021hypergraph,chen2022thinking,wu2023consrec,Ye_2025}. 

Despite recent advances, existing methods essentially treat group decision-making modeling as a simple aggregation of individual user preferences, assuming that a fixed aggregation function can capture the group preferences. This assumption struggles to model and capture the complex dynamics of real-world group decision-making processes. In reality, group decisions are influenced by various social and contextual factors, such as group topics \cite{guo2021hierarchical} and leadership roles \cite{gan2025large}, as well as dynamic interactions between group members \cite{wu2023consrec}. These dynamic interactions often involve preference expression, compromise, and consensus-building for final decisions. As a result, simple aggregation approaches fail to reflect the nuanced and evolving nature of group decision-making behaviors, thereby limiting recommendation performance.

To address these limitations, recent advances in Large Language Model (LLM)-driven agents offer new possibilities for simulating real-world group decision-making \cite{park2023generative}. Integrating LLM-driven agents into GR brings three key advantages \cite{cai2025agentic}. First, their rich world knowledge and strong language understanding capability help to extract deep semantic preferences from the historical interactions of users and groups. Second, their role-playing and text generation abilities allow them to simulate user behaviors such as preference expression and influence propagation during group discussions. Third, their complex reasoning and self-reflection capabilities make them well-suited for modeling negotiation, compromise, and consensus-building in a human-like manner. Leveraging these advantages, we propose a novel agent-based GR paradigm that leverages LLM-driven agents to capture semantic signals and simulate complex group decision-making processes, as shown in Figure~\ref{fig0}.b. Despite their potential, applying LLM-driven agents to GR is non-trivial, which poses two key challenges. First, while LLM-driven agents excel at semantic understanding, they struggle to model high-order graph structural dependencies among users, items, and groups, which are essential for capturing collaborative filtering signals \cite{wang2025unleashing}. Second, current agent-based studies primarily target individual user recommendations, leaving the intricacies of group-level decision-making underexplored, which involves the integration of group topic guidance, leadership influence, multi-user interactions, and consensus-building in a unified framework \cite{jannach2025rethinking}.

To address the aforementioned challenges, we propose a novel Semantic-aware \textbf{\underline{Agent}}ic Group Decision-making Simulator for \textbf{\underline{G}}roup \textbf{\underline{R}}ecommendation, named AgentGR. AgentGR jointly captures collaborative-semantic user preferences for member-role-playing and simulates complex group interactions to reflect real-world group decision-making processes, thereby boosting recommendation performance. To capture collaborative-semantic user preferences for member-role-playing, we introduce a semantic meta-path guided Chain-of-Preference (CoP) reasoning mechanism. This mechanism combines LLMs’ semantic understanding with composite meta-path modeling to extract rich textual features while incorporating high-order graph structural connectivity patterns. In this way, it enables unified preference learning from both semantic and collaborative views. To simulate complex interactions of group decision-making, we first introduce two semantic-aware recognition modules: a group topic recognition module that extracts group-level preferences by aggregating intra-group (group-item view) and inter-group (group-group view) semantic signals, and a leader recognition module that identifies influential members based on their semantic alignment with group topics. Building on these factors, we further devise two multi-agent simulation strategies to model multi-user interactions and consensus-building. A static strategy generates ranked group preferences through a predefined workflow for efficient and controllable recommendations, while a dynamic strategy simulates multi-round interactions among members to model compromise and consensus more accurately. These two strategies offer flexibility to balance efficiency and effectiveness across various application needs. Extensive experiments on two real-world datasets show that AgentGR significantly outperforms state-of-the-art GR methods in both recommendation accuracy and decision-making simulation, confirming its effectiveness.

The main contributions are summarized as follows:
\begin{itemize}
    \item To the best of our knowledge, our work is the first to introduce an agent-based GR paradigm. Based on this, we propose AgentGR, a novel model that integrates semantic information and complex group decision-making modeling through LLM-driven multi-agent simulation.
    \item We design a semantic meta-path guided CoP reasoning mechanism that unifies collaborative filtering signals and semantic knowledge, enabling more accurate user preference profiling and member-role-playing.
    \item To simulate group decision-making, we firstly introduce two semantic recognition modules to model key social and contextual factors, including group topics and leadership. Building on these, we further devise two multi-agent simulation strategies to model group consensus-building and improve recommendation performance.
    \item We conduct extensive experiments on two real-world datasets. Results show that AgentGR significantly outperforms state-of-the-art baselines by at least 9.74\% in HR and NDCG metrics, demonstrating its effectiveness.
\end{itemize}

\section{Related Works}
\subsection{Group Recommendation}
Existing GR methods focus on extracting individual user preferences from user-item interactions and then aggregating them to generate group-level recommendations \cite{yin2019social}. According to the differences in aggregation approaches, these methods can fall into two categories: score aggregation and representation aggregation. Early studies rely on heuristic score aggregation rules, such as average satisfaction \cite{boratto2011state}, most pleasure \cite{baltrunas2010group}, or least misery \cite{amer2009group}. Despite their straightforwardness, these studies are limited by their oversimplification and inflexibility, failing to account for the complex interactions during group decision-making processes. With the advancement of deep learning, later studies utilize advanced deep neural network architectures for aggregating user preference representations rather than recommendation scores. For example, some efforts \cite{yin2019social,vinh2019interact,he2020game,cao2019social,deng2021knowledge} employ attention mechanisms to model user-group influence and enhance group preference representations. Recent efforts \cite{jia2021hypergraph,guo2021hierarchical,chen2022thinking,wu2023consrec} further improve recommendation performance by incorporating advanced graph-based techniques, such as hypergraph neural networks. These GNN-based methods adeptly capture the high-order connectivity among users, groups, and items, significantly enhancing the recommendation performance. However, most existing methods lack the fine-grained modeling of complex and dynamic group decision-making processes, which limits their recommendation effectiveness \cite{Ye_2025}.


\begin{figure*}[t]
\centering
\includegraphics[width=0.95\textwidth]{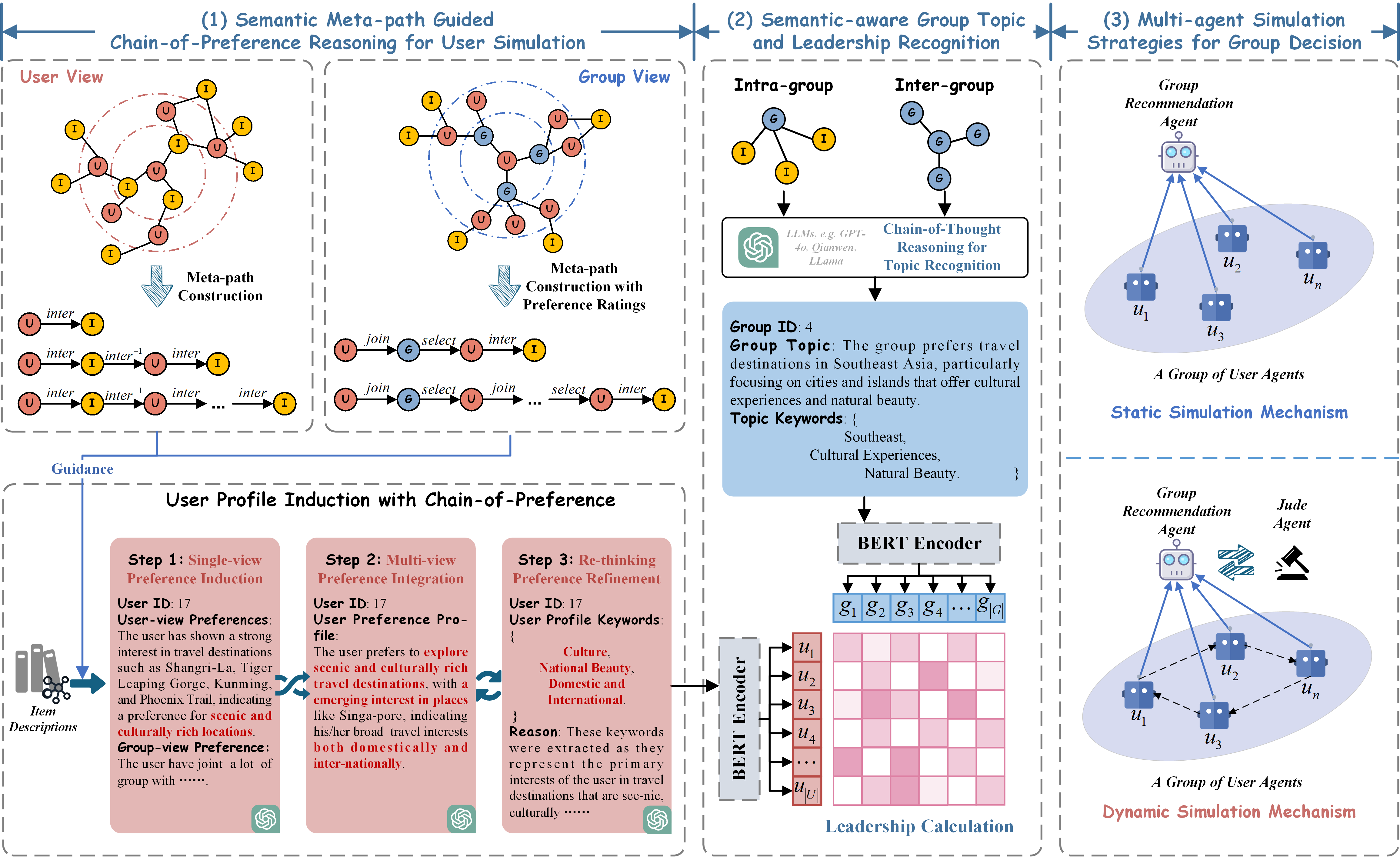} 
\caption{Framework of Semantic-aware \textbf{\underline{Agent}}ic Group Decision-making Simulator for \textbf{\underline{G}}roup \textbf{\underline{R}}ecommendation (AgentGR), which includes three core stages: semantic meta-path guided chain-of-preference reasoning for user simulation, semantic-aware recognition for group topics and leadership, and multi-agent simulation strategies for group decision.}
\label{fig1}
\end{figure*}

\subsection{LLMs for Recommendation}
LLMs \cite{ouyang2022training}, known for their broad world knowledge and strong semantic reasoning capabilities, have been increasingly explored in recommendation systems for feature enhancement \cite{ren2024representation,zhou2025dual} or direct recommendation generation \cite{zhang2025collm,bao2023tallrec,gao2025sprec}. More recently, LLM-based agents have demonstrated impressive abilities in simulating human cognition and behavior, attracting growing interest in the field. For example, Zhang et al. \cite{zhang2025llm} simulate user interaction behaviors with user agents to capture preferences and emotions. Moreover, Zhang et al. \cite{zhang2024agentcf} introduce both user and item agents to emulate user-item interactions for collaborative filtering. Cai et al. \cite{cai2025agentic} propose a feedback loop framework that simulates user-system interaction via user and recommender agents. The aforementioned studies have demonstrated the effectiveness and feasibility of using AI Agents for user simulation. However, these studies primarily focus on individual user simulation. Modeling group decision-making, which involves group topic guidance, leadership influence, multi-user interactions, and consensus-building, remains largely unexplored \cite{jannach2025rethinking}. Moreover, while LLMs are effective at semantic modeling, they still struggle to capture collaborative filtering signals hidden in the high-order connectivity among users, groups, and items \cite{zhu2024collaborative}. In this paper, we strive to break through the challenges faced by LLM-driven agents in collaborative-semantic preference extraction and group decision-making simulation for GR.

\section{Problem Formulation}
In this paper, let $U=\left\{u_1, u_2, \cdots, u_{|U|}\right\}$, $I=\left\{i_1, i_2, \cdots, i_{|I|}\right\}$, and $G=\left\{g_1, g_2, \cdots, g_{|G|}\right\}$ denote the sets of users, items, and groups, respectively, with $|U|$, $|I|$, and $|G|$ representing the total number of users, items, and groups. We use $Y_{|G|\times|I|}=\left\{ y_{gi} | g \in G,i \in I\right\}$ to represent group-item interactions, where $y_{gi}=1$ if group $g$ has interacted with item $i$, and $y_{gi}=0$ otherwise. Similarly, we leverage $X_{|U|\times|I|}=\left\{ x_{ui} | u \in U,i \in I\right\}$ to represent user-item interactions, where $x_{ui}=1$ if user $u$ has interacted with item $i$, and $x_{ui}=0$ otherwise. Moreover, $B_{|G|\times|U|}=\left\{ b_{gu} | g \in G,u \in U\right\}$ is used to represent group-user affiliations, where $b_{gu}=1$ if group $g$ includes user $u$, and $b_{gu}=0$ otherwise. Given a target group $\hat{g} \in G$ and a candidate item set $\hat{I} \subseteq I$, GR aims to identify a ranked list of items from $\hat{I}$ that best match the preferences of group $\hat{g}$.


\section{Methodology}
In this section, we introduce AgentGR in detail. The overall framework of AgentGR is shown in Figure~\ref{fig1}, which mainly includes three core stages: semantic meta-path guided chain-of-preference reasoning for user simulation, semantic-aware recognition for group topics and leadership, and multi-agent simulation strategies for group decision.

\subsection{Semantic Meta-Path Guided Chain-of-Preference Reasoning for User Simulation}
In GR, users, groups, and items form a Heterogeneous Information Network (HIN) with diverse and complex relational structures. These graph structure dependencies contain collaborative filtering signals that are essential for preference learning \cite{wang2019unified,gong2020attentional}. While LLMs excel at extracting semantic features from text, they struggle to extract high-order graph structural dependencies from such HIN \cite{zhu2024collaborative}. To bridge this gap, we introduce meta-paths, which are semantically composite sequences connecting different types of nodes in the HIN, to guide LLMs in focusing on relevant graph structural semantics. Building on this, we propose a semantic meta-path guided Chain-of-Preference (CoP) reasoning mechanism that integrates rich semantic knowledge with high-order collaborative filtering signals for improved preference learning. This mechanism includes two key components: multi-view meta-path construction and user profiling with CoP reasoning.

\subsubsection{Multi-view Meta-path Construction.}
To effectively extract user preferences from behavioral data, we construct two meta-path views, including a user view and a group view, to facilitate preference learning from complementary perspectives \cite{ma2022crosscbr}. In the user view, we design meta-paths of the form $user\stackrel{inter}{\longrightarrow}item\stackrel{{inter}^{-1}}{\longrightarrow}user\stackrel{inter}{\longrightarrow}\cdots\stackrel{inter}{\longrightarrow}item$, to model high-order user preferences from user-item interactions. These meta-paths reflect composite semantic relationships formed by shared interaction histories. Let $A_{UI} \in {\left\{ 0,1 \right\}}^{|U|\times|I|}$ denote the adjacency matrix of user-item interactions $X_{|U|\times|I|}$, the user-view meta-paths can be constructed as:
\begin{equation}
    MP_{UI}^h = {( A_{UI} \circ {A_{UI}}^T )}^{h-1} \circ A_{UI},
\end{equation}
where $\circ$ denotes matrix multiplication, $T$ indicates matrix transposition operation, and $MP_{UI}^h$ represents the meta-path matrix at order $h$ in the user-view. When $h=1$, the meta-path matrix reduces to $MP_{UI}^1 = A_{UI}$, representing the original user-item interactions. Ultimately, the complete user-view meta-path set is denoted as $\mathcal{MP}_{UI}=\left\{MP_{UI}^1,MP_{UI}^2,\cdots,MP_{UI}^H \right\}$, with $H$ as the maximum order of meta-paths.

In the group view, constructing meta-paths over all group members often introduces noise due to group size and user diversity. To mitigate this, we adopt an LLM-based adaptive member selection strategy that identifies members whose preferences align with group-level interests. In this way, the group-view meta-paths are designed as the form $user\stackrel{join}{\longrightarrow}$
$group\stackrel{select}{\longrightarrow}user\stackrel{join}{\longrightarrow}\cdots\stackrel{select}{\longrightarrow}user\stackrel{inter}{\longrightarrow}item$. Specifically, for each group $g$, we calculate the preference alignment score $L_{gu}$ between the group and each of its members $u \in \left\{ u|b_{gu}=1 \right\}$ using a three-level semantic rating function $f_{rate}(\cdot)$ driven by LLMs \footnote{The LLM utilized in AgentGR is optional. In our implementation, we use GPT-4o. Due to space limitations, all the detailed prompts are provided in the code implementation.}:
\begin{equation}
    L_{gu}=f_{rate}(\Gamma_g,\Gamma_u),\space L_{gu} \in \left\{ high,medium,low \right\},
\end{equation}
where $\Gamma_g=\left\{ \tau_1, \cdots, \tau_j, \cdots, \tau_{|\Gamma_g|}\right\}$ and $\Gamma_u=\left\{ \tau_1, \cdots, \tau_p, \cdots, \tau_{|\Gamma_u|}\right\}$ represent the sets of textual descriptions of items previously interacted with by the group $g$ and member user $u$, respectively. We retain only the members rated as $high$, and construct a filtered group-user adjacency matrix $\hat{A}_{GU} \in {\left\{ 0,1 \right\}}^{|G|\times|U|}$, where $\hat{A}_{gu}=1$ if $L_{gu}=high$, and 0 otherwise. Let $A_{UG} \in {\left\{ 0,1 \right\}}^{|U|\times|G|}$ denote the transpose of the adjacency matrix of group-user affiliations $B_{|G|\times|U|}$. Then, the group-view meta-path $MP_{UGI}^h$ of order $h$ is defined as:
\begin{equation}
    MP_{UGI}^h = {( A_{UG} \circ {\hat{A}_{GU}} )}^{h-1} \circ A_{UI}.
\end{equation}
$\mathcal{MP}_{UGI}=\left\{MP_{UGI}^2,MP_{UGI}^3,\cdots,MP_{UGI}^H \right\}$ denotes the group-view meta-path set. By combining user-view and group-view meta-paths, our multi-view approach captures high-order, semantically meaningful connectivity patterns. This guides LLMs to better model collaborative filtering signals and semantic features for preference learning.

\subsubsection{User Profiling with CoP Reasoning.}
To effectively capture implicit user preferences from multi-view interaction data, we introduce a CoP reasoning mechanism, inspired by the Chain-of-Thought reasoning \cite{wei2022chain}. This stepwise process mimics human inductive and reflective reasoning, enabling LLMs to infer user preferences in a more fine-grained manner. It captures both semantic and collaborative filtering signals through three stages: single-view preference induction, multi-view preference integration, and re-thinking preference refinement. 

In the single-view preference induction stage, LLMs are prompted to extract user preferences separately from the user and group views with the guidance of high-order meta-paths. This process yields natural language descriptions of user preferences for each view:
\begin{equation}
    E_{UI}^{pre}=f_{pre}(\mathcal{MP}_{UI}), E_{UGI}^{pre}=f_{pre}(\mathcal{MP}_{UGI}),
\end{equation}
where $E_{UI}^{pre}$ and $E_{UGI}^{pre}$ represent the user-view and group-view user preferences, respectively. $f_{pre}(\cdot)$ denotes the LLM-based preference extraction function.

In the multi-view preference integration stage, LLMs are prompted to fuse single-view preference descriptions into a unified user preference profile:
\begin{equation}
    E_U^{pre}=f_{agg}^{pre}(E_{UI}^{pre},E_{UGI}^{pre}),
\end{equation}
where $E_U^{pre}$ denotes the integrated user profiles in textual form, and $f_{agg}^{pre}(\cdot)$ represents the LLM-based aggregation function. Notably, to mitigate noise from high-order paths, the prompts emphasize direct user-item interactions while treating higher-order signals as auxiliary.

In the re-thinking preference refinement stage, to avoid token overload and the ``lost-in-the-middle" issue \cite{wu2024efficient} during later group decision simulation, LLMs are prompted to re-think the integrated preference profiles and distill them into concise, informative keywords:
\begin{equation}
    E_U^{key}=f_{key}(E_U^{pre}),
\end{equation}
where $E_U^{key}$ denotes the refined user preference profiles. And $f_{key}(\cdot)$ extracts representative keywords and provides justifications to enhance the robustness and interpretability of preference refinement. Finally, these refined user preference profiles are used to guide LLM-driven agents in simulating individual user behaviors, laying the groundwork for accurate group decision modeling.
\subsection{Semantic-aware Group Topic and Leadership Recognition}

\subsubsection{Group Topic Recognition.}
Groups often form around shared interest topics, but group-item interactions (intra-group) are typically sparse, making it difficult to infer topics from intra-group data alone \cite{guo2021hierarchical,gan2025large,wu2023consrec}. To address this, we incorporate both intra- and inter-group (group-group similarity) signals for topic recognition. Specifically, for a group $g$, we first prompt LLMs to summarize its topic $e_g^{intra}$ based on the descriptions of its interacted items:
\begin{equation}
    e_g^{intra}=f_{topic}^{intra}(\Gamma_g),
\end{equation}
where $\Gamma_g$ denotes description texts of items interacted by $g$, and $f_{topic}^{intra}(\cdot)$ is the LLM-based intra-group topic extraction function. Next, we identify a set of similar groups $N_g$ that share overlapping item interactions with group $g$. Their intra-group topics are aggregated to derive the inter-group topic $e_g^{inter}$ of group $g$:
\begin{equation}
    e_g^{inter}=f_{topic}^{inter}(\sum_{g_q \in N_g}e_{g_q}^{intra}),
\end{equation}
where $f_{topic}^{inter}(\cdot)$ is the LLM-based inter-group topic extraction function used to abstract common topics across related groups. Finally, we integrate both intra- and inter-group topics to form a complete topic description of group $g$:
\begin{equation}
    e_g^{topic}=f_{agg}^{topic}(e_g^{intra}, e_g^{inter}),
\end{equation}
where $f_{agg}^{topic}(\cdot)$ fuses the two kinds of topics with emphasis on intra-group content. The set of group topic descriptions for all groups is denoted by $E_G^{topic}$. The joint modeling of intra- and inter-group semantics enables a more robust and context-aware understanding of group-level preferences.

\subsubsection{Group Leader Recognition.}
Leadership also plays a vital role in group choices, as decisions often align with those of influential members \cite{gan2025large}. To identify group leaders, we assess the semantic alignment between each member's preference profile and the group topic. Rather than relying on costly LLM-based semantic comparisons, we adopt a pre-trained BERT \cite{devlin2019bert}encoder to embed both the group topic and each member's refined preference profile:
\begin{equation}
    v_g=BERT(e_g^{topic}),v_u=BERT(e_u^{key}),
\end{equation}
where $BERT(\cdot)$ denotes the BERT encoding operation. $v_g$ and $v_u$ are the semantic embeddings of the group topic and user preference profile, respectively. We then select the member with the highest cosine similarity as the leader:
\begin{equation}
    u_{leader}=\mathop{\arg\max}\limits_{u\in U_g}\space cos(v_g,v_u),
\end{equation}
where $cos(\cdot)$ computes cosine similarity and the user $u_{leader}$ is identified as the leader of the group $g$. This efficient strategy captures semantic alignment while maintaining scalability across large GR datasets.

\subsection{Multi-agent Simulation Strategies for Group Decision-Making}
To better reflect real-world group decision-making, we design two multi-agent simulation strategies, static and dynamic, drawing on the role-playing and interaction capabilities of LLM-driven agents. These strategies both leverage the previously identified group-level factors, including the group topics and leaders.
\subsubsection{Static Simulation Strategy.}
The static simulation strategy adopts a predefined two-stage workflow pipeline: user-level ranking followed by group-level re-ranking, which involves multiple user agents and one group recommendation agent. Firstly, each user agent individually ranks candidate items $\hat{I}$ based on her/his preference profile and the group topic:
\begin{equation}
    \hat{I}_u=rank(e_u^{key},e_{\hat{g}}^{topic},\hat{I}),\space e_u^{key} \in E_U^{key},
\end{equation}
where $\hat{I}_u \subseteq \hat{I}$ is the ranked item list generated by agent of user $u$ in the target group $\hat{g}$, and $rank(\cdot)$ denotes the item ranking operation of user agents. Next, the group agent re-ranks and generates the final recommendations by integrating all individual rankings and considering the influence of the group leader:
\begin{equation}
    \hat{I}_{\hat{g}}^s=re\_rank(e_{\hat{g}}^{topic},\left\{\hat{I}_1, \cdots, \hat{I}_u, \cdots, \hat{I}_{|U_{\hat{g}}|} \right\},\hat{I},u_{leader}),
\end{equation}
where $\hat{I}_{\hat{g}}^s$ represents the final recommendations from the static simulation strategy. This pipeline design ensures controllability and efficiency in multi-agent decision-making modeling.

\subsubsection{Dynamic Simulation Strategy.}
To better capture the nuances of human group decision-making, the dynamic simulation strategy allows open-ended discussion among agents. Inspired by AutoGen \cite{wu2024autogen}, user agents engage in multi-round discussions, guided by the group topic and leader. Subsequently, a group agent summarizes the discussions and yields a recommendation result, while an external judgment agent evaluates whether consensus is achieved. If consensus is reached, the process concludes. Otherwise, additional discussion rounds begin. The process is defined as:
\begin{equation}
    \hat{I}_{\hat{g}}^d=dy\_rank(e_{\hat{g}}^{topic},\left\{e_1^{key}, \cdots, e_u^{key}, \cdots, e_{U_{\hat{g}}}^{key}\right\},\hat{I},u_{leader}),
\end{equation}
where $\hat{I}_{\hat{g}}^d$ denotes the recommendation result from the dynamic simulation strategy, and $dy\_rank(\cdot)$ represents the iterative, multi-round dynamic reasoning procedure among agents. This dynamic strategy models compromise and opinion evolution, producing more realistic group decisions.

Both simulation strategies can be flexibly chosen based on the application requirements: the static strategy focuses on efficiency, while the dynamic strategy emphasizes fidelity to real-world interactions.
\section{Experiments}
To evaluate the effectiveness of our proposed AgentGR, we conduct extensive experiments on two publicly available datasets for the GR task, aiming to address the following research questions: \textbf{RQ1}: How does AgentGR perform compared with the state-of-the-art group recommendation methods? \textbf{RQ2}: What are the performance and costs of static and dynamic simulation strategies? \textbf{RQ3}: How do the novel components of AgentGR affect the overall recommendation performance? \textbf{RQ4}: Can AgentGR effectively understand and extract the preferences of users? \textbf{RQ5}: How do different scales and types of LLMs affect the performance of AgentGR? \textbf{RQ6}: Can AgentGR provide an intuitive impression for group decision-making? \textbf{RQ7}: How do the key hyperparameters of AgentGR affect the overall recommendation performance?

\subsection{Experimental Setup}

\begin{table}[h]
\caption{Statistics of GR datasets, where U-I and G-I denote user-item and group-item interactions, respectively.}
\centering
\begin{tabular}{c|ccccc}
    \hline
    Dataset & \#Users & \#Groups & \#Items & \#U-I & \#G-I \\
    \hline
    MafengwoS & 11,027 & 1,215 & 1,236 & 6,563 & 1,886 \\
    Weeplaces & 290 & 298 & 7,829 & 15,070 & 697 \\
    \hline
\end{tabular}
\label{table1}
\end{table}

\begin{table*}[t]
\caption{Overall performance comparison of all methods on the GR task. “Improv." indicates the relative improvement of our AgentGR (\textbf{bold}) compared with the sub-optimal method (\underline{underline}). Here, AgentGR employs the static simulation strategy.}
\centering
\begin{tabular}{c|c|cccccccc|cc}
    \hline
    Dataset & Metric & AGREE & GroupIM & HCR & CubeRec & ConsRec & AlignGroup & LLM4GR & DisRec & AgentGR & Improv. \\
    \hline
    \multirow{4}{*}{MafengwoS} & HR@5 & 0.4021 & 0.3643 & 0.4166 & 0.6770 & 0.4949 & 0.6873 & 0.7040 & \underline{0.7526} & \textbf{0.8881} & 18.00\% \\
    & HR@10 & 0.5533 & 0.5653 & 0.6025 & 0.7766 & 0.6478 & 0.7440 & 0.7642 & \underline{0.8196} & \textbf{0.9260} & 12.98\% \\
    & NDCG@5 & 0.2623 & 0.2316 & 0.4321 & 0.5540 & 0.3788 & 0.5770 & 0.5587 & \underline{0.6626} & \textbf{0.7760} & 17.11\% \\
    & NDCG@10 & 0.3115 & 0.2973 & 0.5169 & 0.5867 & 0.4286 & 0.5959 & 0.5812 & \underline{0.6842} & \textbf{0.7913} & 15.65\% \\
    \hline
    \multirow{4}{*}{Weeplaces} & HR@5 & 0.4089 & 0.3826 & 0.3423 & 0.3456 & 0.4027 & 0.4195 & 0.4060 & \underline{0.4295} & \textbf{0.5067} & 17.97\% \\
    & HR@10 & 0.5550 & 0.4664 & 0.3926 & 0.4362 & \underline{0.5705} & 0.5638 & 0.5604 & 0.5403 & \textbf{0.6644} & 16.46\% \\
    & NDCG@5 & 0.2906 & 0.2960 & 0.2933 & 0.2726 & 0.2943 & 0.3155 & 0.3114 & \underline{0.3397} & \textbf{0.3728} & 9.74\% \\
    & NDCG@10 & 0.3378 & 0.3209 & 0.3090 & 0.3017 & 0.3491 & 0.3622 & 0.3632 & \underline{0.3749} & \textbf{0.4246} & 13.26\% \\
    \hline
\end{tabular}
\label{table2}
\end{table*}

\subsubsection{Datasets.}
We conduct experiments on two real-world public datasets: MafengwoS\footnote{https://github.com/FDUDSDE/WWW2023ConsRec} and Weeplaces\footnote{https://stephenliu0423.github.io/datasets.html}. MafengwoS is collected from a tourism website where users create or join group travel activities. Weeplaces is a typical check-in dataset for group gathering activities in various restaurants. Unlike other existing GR datasets, only these two datasets contain sufficiently rich textual data that can support our study. Their statistics are shown in Table \ref{table1}. To ensure fair comparison, we adopt the leave-one-out strategy following the standard GR protocol used by all baselines officially \cite{10.1145/3209978.3209998,jia2021hypergraph,wu2023consrec,xu2024aligngroup,Ye_2025}, designating the last interactions as the test set and the remaining interactions as the training set. Additionally, to reduce computational load, we follow prior research \cite{10.1145/3209978.3209998,wu2023consrec,xu2024aligngroup} by randomly selecting 50 negative samples for each positive sample in the test set. Notably, AgentGR constructs user and group profiles exclusively from the training set, thereby strictly avoiding any test data leakage.

\subsubsection{Baselines.}
To evaluate the effectiveness of AgentGR, we select eight state-of-the-art GR baselines for comparison, including: AGREE, GroupIM, HCR, CubeRec, ConsRec, AlignGroup, DisRec, and LLM4GR. More details are provided below:
\begin{itemize}
    \item \textbf{AGREE} \cite{10.1145/3209978.3209998} leverages attention networks and neural collaborative filtering to learn aggregation strategies for member preference integration in group recommendations, thereby alleviating the cold-start issues.
    \item \textbf{GroupIM} \cite{sankar2020groupim} maximizes the mutual information between users and groups by utilizing contextual preference weighting to address the sparsity of group interactions in ephemeral group recommendations.
    \item \textbf{HCR} \cite{jia2021hypergraph} proposed a dual-channel hypergraph convolutional network that captures cross-group user-item collaboration via a member-level hypergraph and models group-wide preferences using a similarity-based group-level graph.
    \item \textbf{CubeRec} \cite{chen2022thinking} employs a hypercube vector space to represent group preferences, introducing a novel distance metric and utilizing self-supervised learning to improve recommendation performance.
    \item \textbf{ConsRec} \cite{wu2023consrec} designs member-level, item-level, and group-level views to capture group consensus, employing a hypergraph neural network for member-level aggregation.
    \item \textbf{AlignGroup} \cite{xu2024aligngroup} captures group consensus by modeling both intra- and inter-group relations using a hypergraph neural network and employing self-supervised alignment to coordinate consensus with member preferences.
    \item \textbf{DisRec} \cite{Ye_2025} disentangles user preferences from social influence to address preference bias in group recommendations, introducing social-based contrastive learning to mitigate data sparsity.
    \item \textbf{LLM4GR} \cite{tommasel2024fairness} develops an evaluation framework to examine the impact of sensitive attributes on group recommendations generated by LLMs, revealing their interaction patterns. We extend this method to provide group recommendations.
\end{itemize}

\subsubsection{Evaluation Metrics.}
Following prior work \cite{wu2023consrec,Ye_2025}, we adopt HR@K and NDCG@K as the evaluation metrics, setting K=\{5,10\} to measure the effectiveness of Top-K recommendation results. Higher metric values correspond to better recommendation performance.

\subsubsection{Implementation Details.}
We implemented the AgentGR using non-distributed training in Python 3.8.19 and PyTorch 2.3.0. All experiments were conducted on a Linux machine configured with two 4090 GPUs. We selected GPT-4o as our base LLM and used the OpenAI API, without fine-tuning applied. All experimental indicators of AgentGR in this paper adopt the static simulation strategy. For all baselines, we use the official implementations released by the authors on GitHub. To ensure fairness, the hyperparameters of all the baselines are extensively tuned within the officially defined hyperparameter spaces.
Due to space limitations, the detailed prompts of AgentGR are provided in our GitHub repository.

\subsection{Overall Performance (RQ1)}
To verify the superiority and effectiveness of our proposed AgentGR, we compare its recommendation performance with that of all compared baselines. The experimental results on the GR task are shown in Table \ref{table2}. Here, the optimal results are highlighted in \textbf{bold}, and the suboptimal results are marked with \underline{underline}. By analyzing the results, we can draw the following insights.

First, AgentGR consistently outperforms all baselines across all metrics. Compared to the sub-optimal baseline, AgentGR achieves at least a 9.74\% improvement on both HR@K and NDCG@K. This is because AgentGR not only integrates textual semantics with high-order collaborative filtering signals for user preference profiling, but also models decision-influencing factors and group interaction dynamics at a fine-grained level. Second, LLM-based methods generally outperform traditional DL-based methods (except for DisRec), highlighting the importance of incorporating textual semantic modeling in GR. Among LLM-based methods, AgentGR achieves an average improvement of 22.67\% on HR@K and 26.95\% on NDCG@K over LLM4GR. This is because LLM4GR only incorporates textual semantic features, whereas AgentGR further captures high-order collaborative signals and simulates group decision-making processes through multi-agent reasoning. In addition, multi-view methods (ConsRec, AlignGroup, and DisRec) yield better performance than single-view methods (AGREE, GroupIM, and HCR), which aligns with prior findings \cite{xu2024aligngroup,Ye_2025} and further validates the reliability of our comparison experiments.

\subsection{Static vs. Dynamic Simulation (RQ2)}

\subsubsection{Performance Analysis.}
\begin{figure}[h]
\centering
\includegraphics[width=\columnwidth]{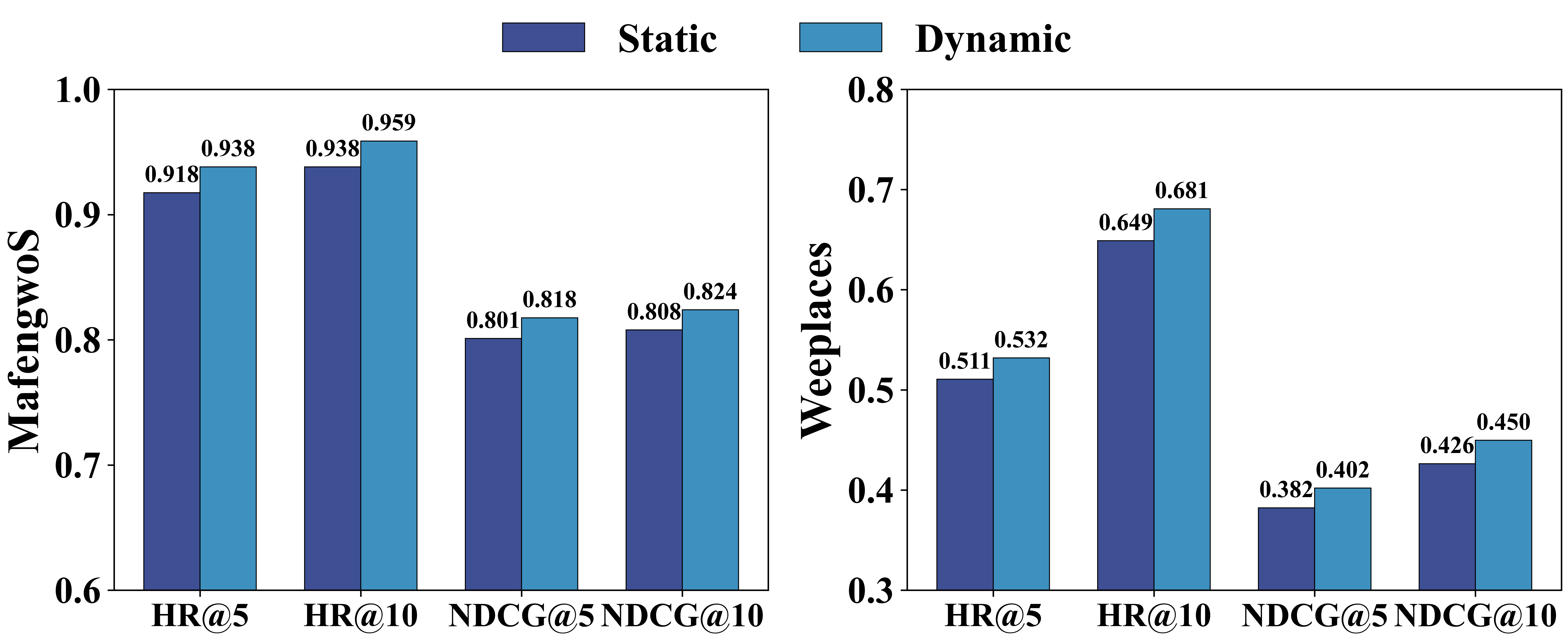} 
\caption{Comparative Experiments between static and dynamic simulation strategies.}
\label{fig2}
\end{figure}

To evaluate the effectiveness of static and dynamic simulation strategies, we conduct a comparative analysis of both strategies. Due to the high cost of dynamic simulation, we randomly select 100 groups from both the MafengwoS and Weeplaces datasets. As shown in Figure~\ref{fig2}, results indicate that both strategies show good performance and the dynamic strategy consistently outperforms the static strategy in recommendation quality. This improvement stems from their fundamental design differences. The static strategy follows a fixed workflow pipeline, offering efficiency and controllability, but fails to capture the interaction and compromise among group members. In contrast, the dynamic strategy supports open-ended user discussions, enabling realistic modeling of compromise and consensus-building, which leads to better recommendations. In practice, each strategy has its strengths: the static strategy suits time-sensitive scenarios, while the dynamic strategy, though more costly, is preferable when recommendation quality is the priority.


\begin{figure}[h]
\centering
\includegraphics[width=\columnwidth]{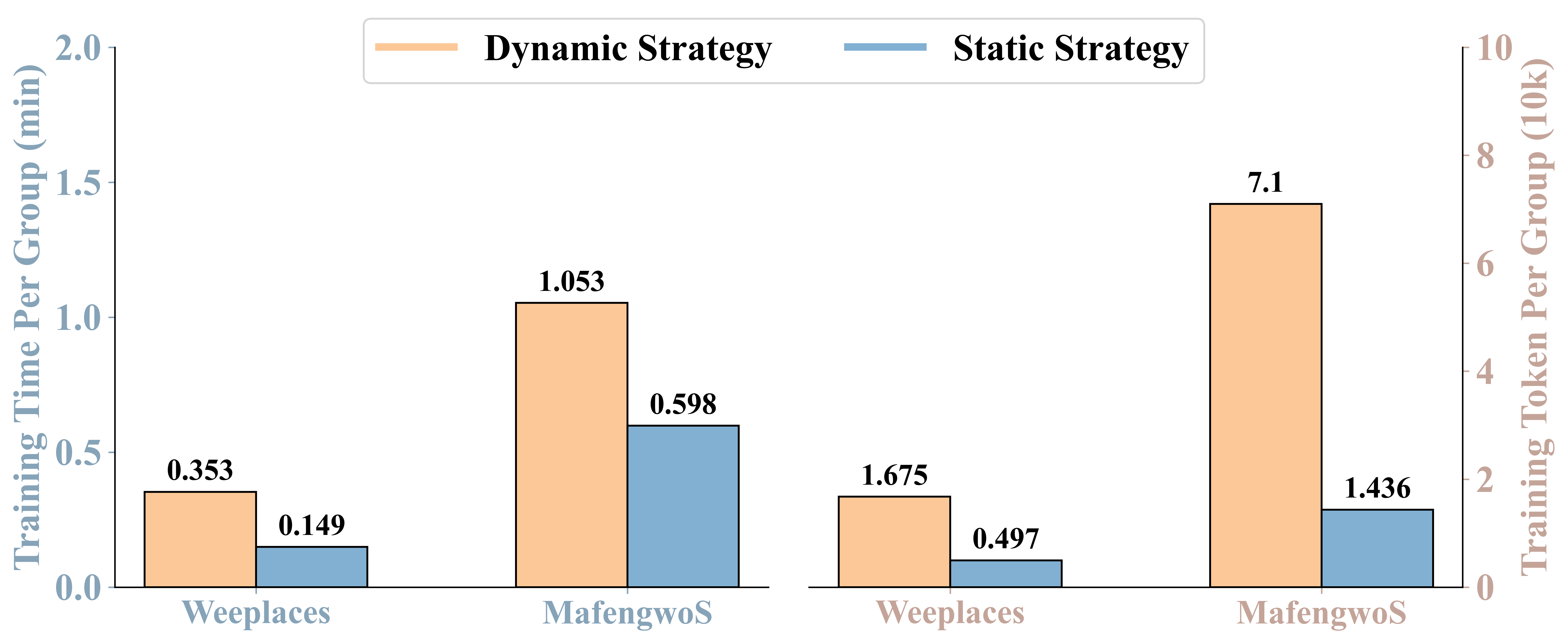} 
\caption{Time and Token cost.}
\label{fig4}
\end{figure}

\subsubsection{Time and Token Analysis.}
In this section, we evaluate the efficiency of static and dynamic simulation strategies by comparing their inference time and token costs, as shown in Figure~\ref{fig4}. Although dynamic simulation strategy yields higher recommendation accuracy, its inference time and token consumption are much higher than those of static simulation strategies. The choice between strategies depends on application needs. The static simulation strategy offers better efficiency and controllability, making it suitable for time- and resource-sensitive scenarios. In contrast, the dynamic simulation strategy captures more detailed multi-user interactions during the group decision-making processes, which benefits tasks where accuracy is the primary concern.

\begin{table}
\caption{Performance comparison of the ablation study. The optimal results are highlighted in \textbf{bold}, suboptimal results are \underline{underlined}, and the worst results are marked with “*".}
\label{table3}
\begin{tabular}{c|cc|cc}
    \hline
    \multirow{2}{*}{Mathod} & \multicolumn{2}{c|}{MafengwoS} & \multicolumn{2}{c}{Weeplaces} \\
    & HR@10 & NDCG@10 & HR@10 & NDCG@10 \\
    \hline
    w/o. M & 0.8726* & 0.7088* & 0.5503* & 0.3566* \\
    w/o. G & 0.9088 & 0.7564 & 0.6001 & 0.3653 \\
    w/o. L & \underline{0.9249} & \underline{0.7848} & \underline{0.6577} & \underline{0.4104} \\
    w/o. S & 0.9208 & 0.7648 & 0.6175 & 0.4056 \\
    \hline
    AgentGR & \textbf{0.9260} & \textbf{0.7913} & \textbf{0.6644} & \textbf{0.4246} \\
    \hline
\end{tabular}
\end{table}

\subsection{Ablation Study (RQ3)}
To assess the contribution of AgentGR's core components, we perform ablation studies by removing each component individually while keeping the others unchanged, yielding four variants: ``w/o. M", ``w/o. G", ``w/o. L", and ``w/o. S" represent the variants of removing semantic meta-path guided CoP reasoning, group topic recognition, group leader recognition, and multi-agent simulation strategies, respectively. Due to space limitations, Table \ref{table3} only shows the results on HR@10 and NDCG@10 across two datasets. The results indicate that removing any component leads to a significant drop in recommendation performance across two datasets, highlighting the indispensable role of each component. Notably, ``w/o. M" shows the most significant decline, underscoring the effectiveness of semantic meta-path guided CoP reasoning in capturing high-order collaborative and semantic signals. In addition, the relatively smaller impact of ``w/o. L" suggests that not all groups exhibit strong leadership, consistent with prior findings \cite{gan2025large}. We will explore adaptive leader modeling in our future research. Moreover, the performance decline of ``w/o. S" further emphasizes the necessity and effectiveness of simulating realistic group decision-making via multi-agent cooperation.

\begin{figure}[h]
    \centering
    \begin{subfigure}[b]{0.85\columnwidth}
        \centering
        \includegraphics[width=\linewidth]{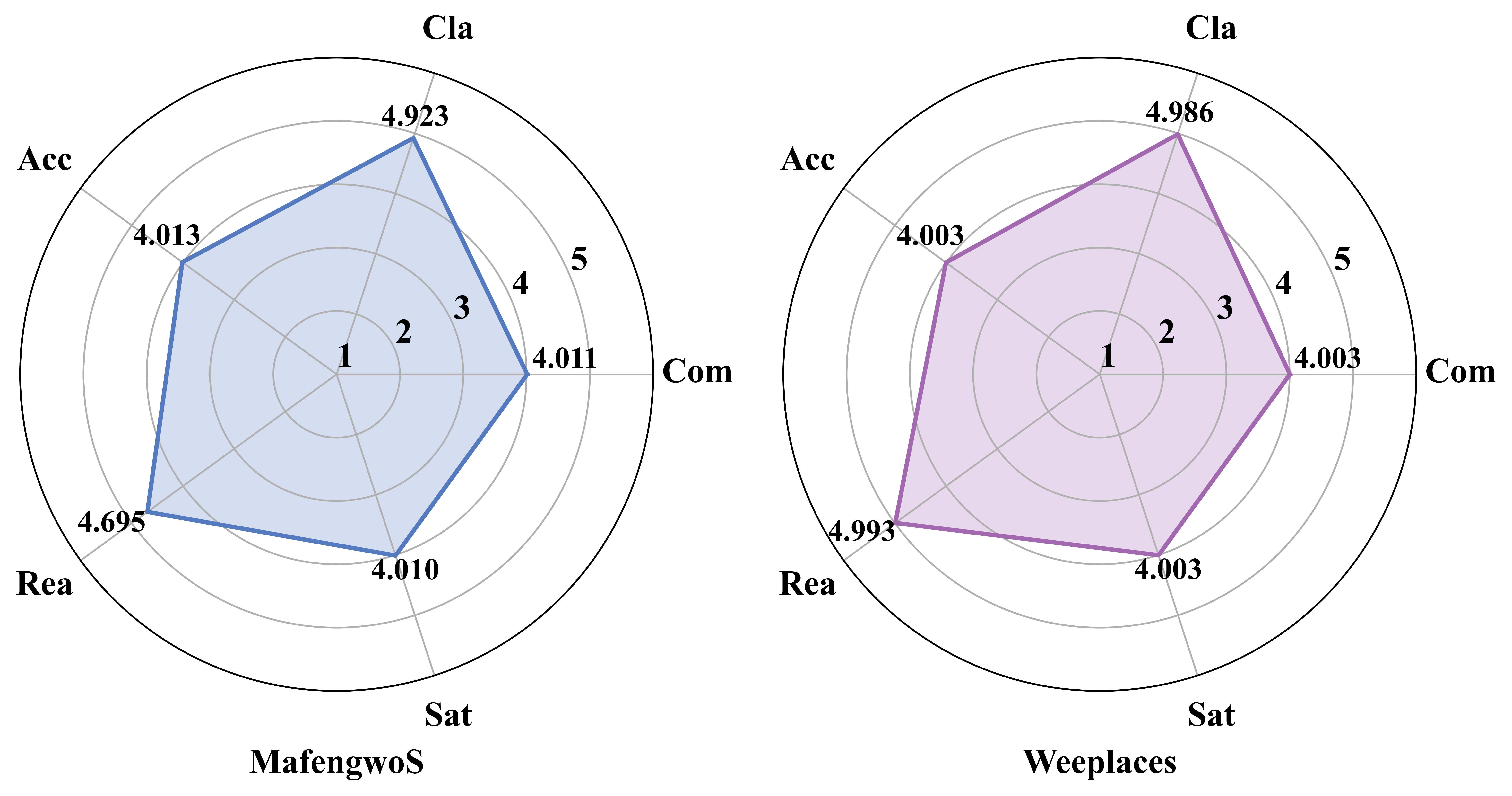}
        \caption{LLM-as-Judge Evaluation} 
        \label{fig6a}
    \end{subfigure}
    \hfill
    \begin{subfigure}[b]{0.9\columnwidth}
        \centering
        \includegraphics[width=\linewidth]{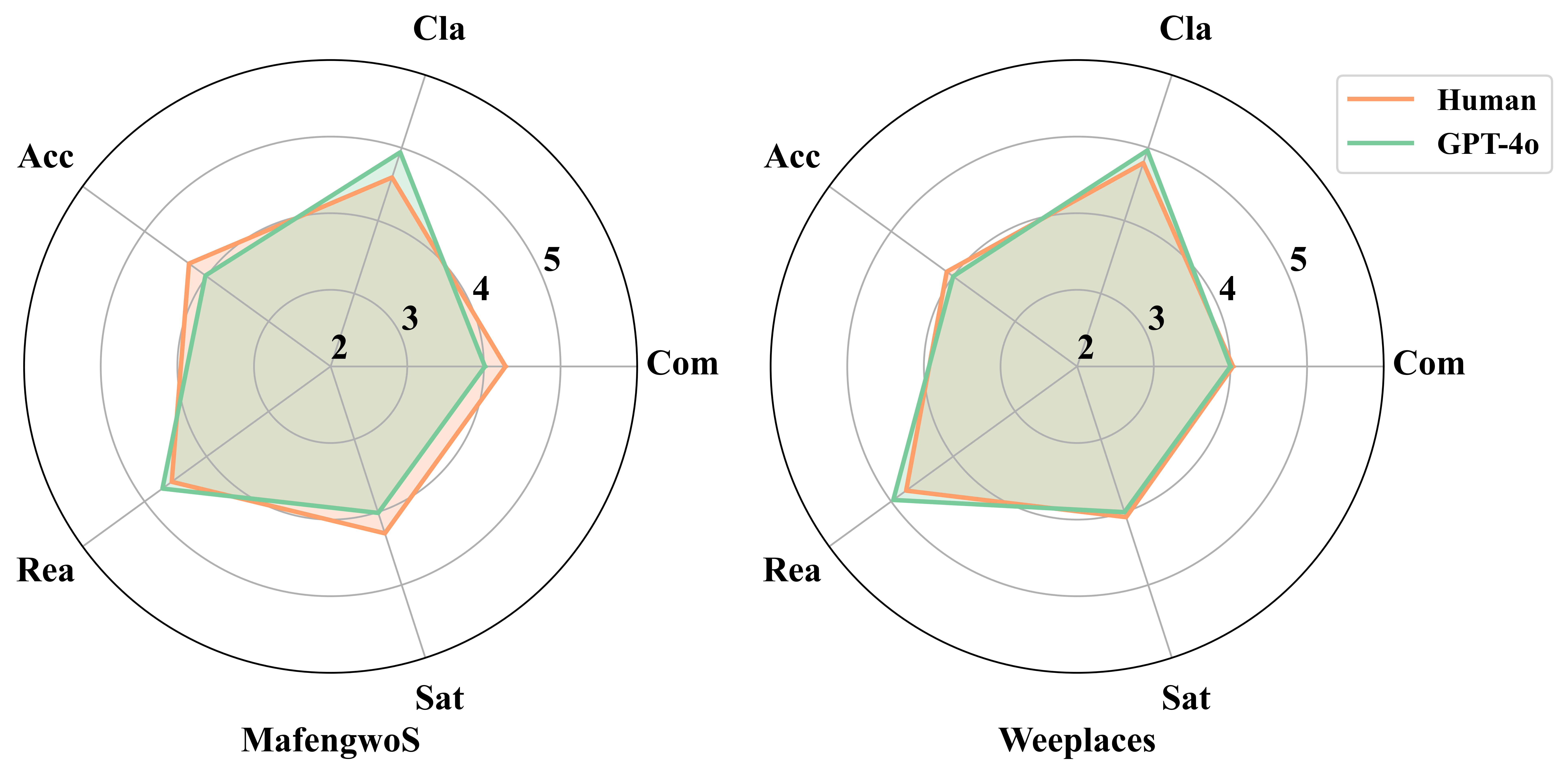}
        \caption{LLM-as-Judge vs. Human-Judgement} 
        \label{fig6b}
    \end{subfigure}
    \caption{Quality Analysis of User Profiling.}
\end{figure}

\subsection{Quality Analysis of User Profiles (RQ4)}
We further examine whether AgentGR can effectively understand and extract user preferences, which is a critical step for reliable group decision simulation. However, evaluating user preference profiling is challenging because explicit ground truth labels are unavailable. Motivated by recent LLM-as-Judge studies \cite{dong2024can,jiang2025beyond,wang2024user,zhouknowledge}, we adopt a qualitative evaluation protocol that uses LLMs as evaluators. Specifically, we employ prompt-driven GPT-4o and define five evaluation metrics: Comprehensiveness (Com), Clarity (Cla), Accuracy (Acc), Reasonableness (Rea), and Satisfaction (Sat). Each metric is scored on a five-point scale, where higher values indicate better performance. Due to space constraints, detailed evaluation prompts are provided in our anonymous GitHub repository. In addition, following prior work \cite{zhang2024large}, we randomly sample 10 percent of the evaluated cases and invite two human annotators to assess the user preference profiles using the same criteria. As shown in Figure 5, AgentGR achieves consistently high scores across all dimensions, indicating its ability to produce meaningful and faithful user preference profiles. Moreover, the strong agreement between GPT-4o and human judgments supports the effectiveness of AgentGR in user preference profiling and validates the reliability of LLM-based evaluation in this setting.

\begin{figure}[h]
\centering
\includegraphics[width=\columnwidth]{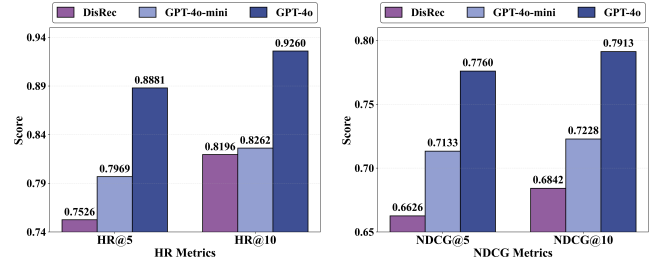} 
\caption{Impact of different LLM scales.}
\end{figure}

\begin{figure}[h]
\centering
\includegraphics[width=0.9\columnwidth]{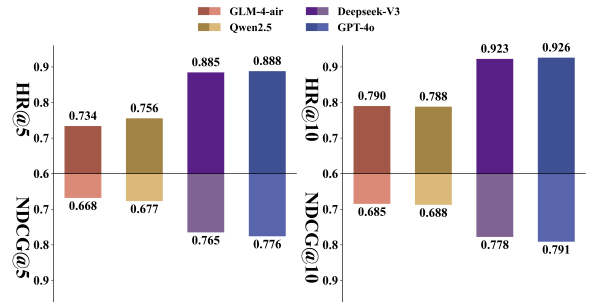} 
\caption{Impact of different LLM types.}
\end{figure}

\begin{figure*}[t]
\centering
\includegraphics[width=0.89\textwidth]{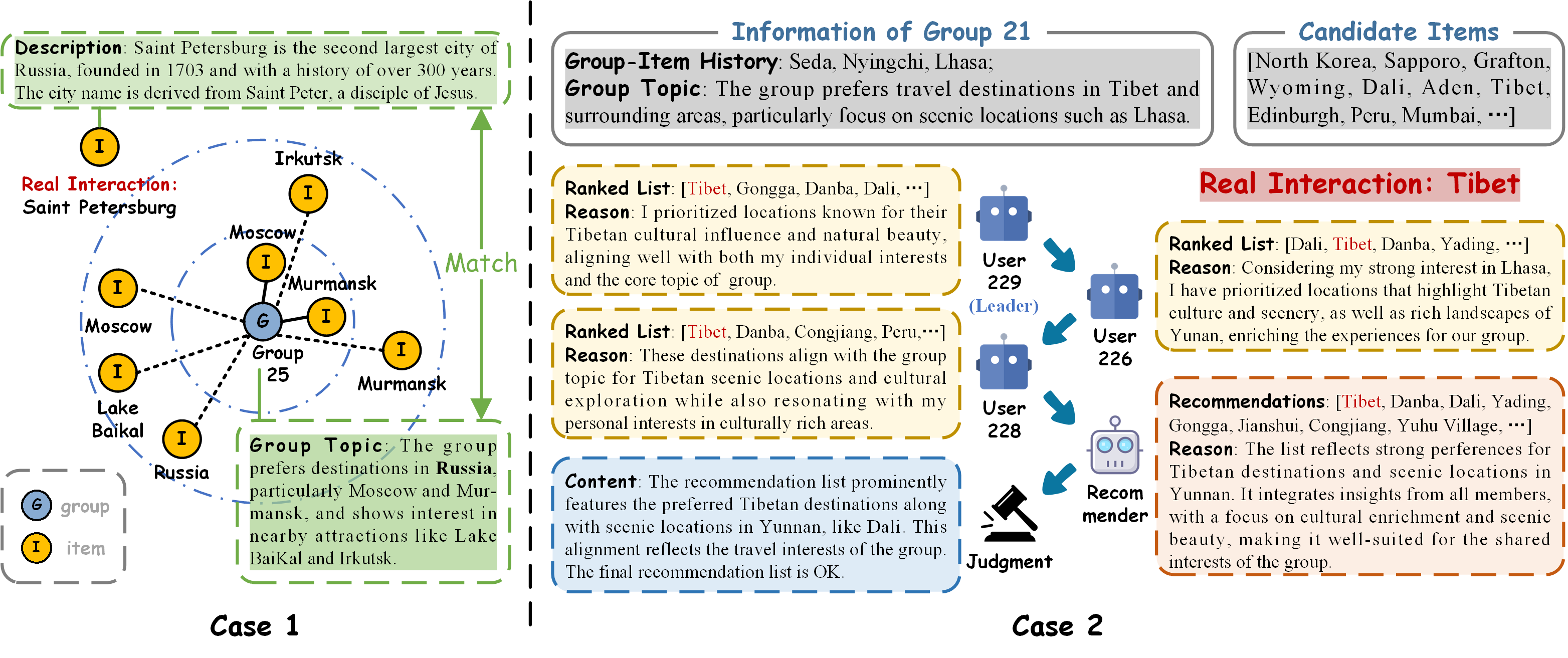} 
\caption{Two real cases in MafengwoS dataset.}
\label{fig5}
\end{figure*}
\subsection{Analysis of LLMs' Scales and Types (RQ5)}
\subsubsection{Impact of LLM Scales.}
We study the effect of LLM scales on AgentGR' performance by comparing two variants that apply GPT-4o (large-scale) and GPT-4o-mini (lightweight), respectively. Figure 5 shows that both variants consistently outperform the optimal conventional GR baseline (DisRec) and larger LLM lead to better recommendation performance. However, compared with the variant based on GPT-4o-mini, the performance gain from GPT-4o is modest relative to its much larger parameter scale, indicating diminishing marginal gains of LLM scaling. These results highlight a practical trade-off between accuracy and efficiency. In resource-limited settings, AgentGR can adopt lightweight LLMs or hybrid strategies, such as using larger LLMs only for group decision simulation while applying smaller LLMs to other modules.

\begin{figure}[b]
\centering
\includegraphics[width=0.95\columnwidth]{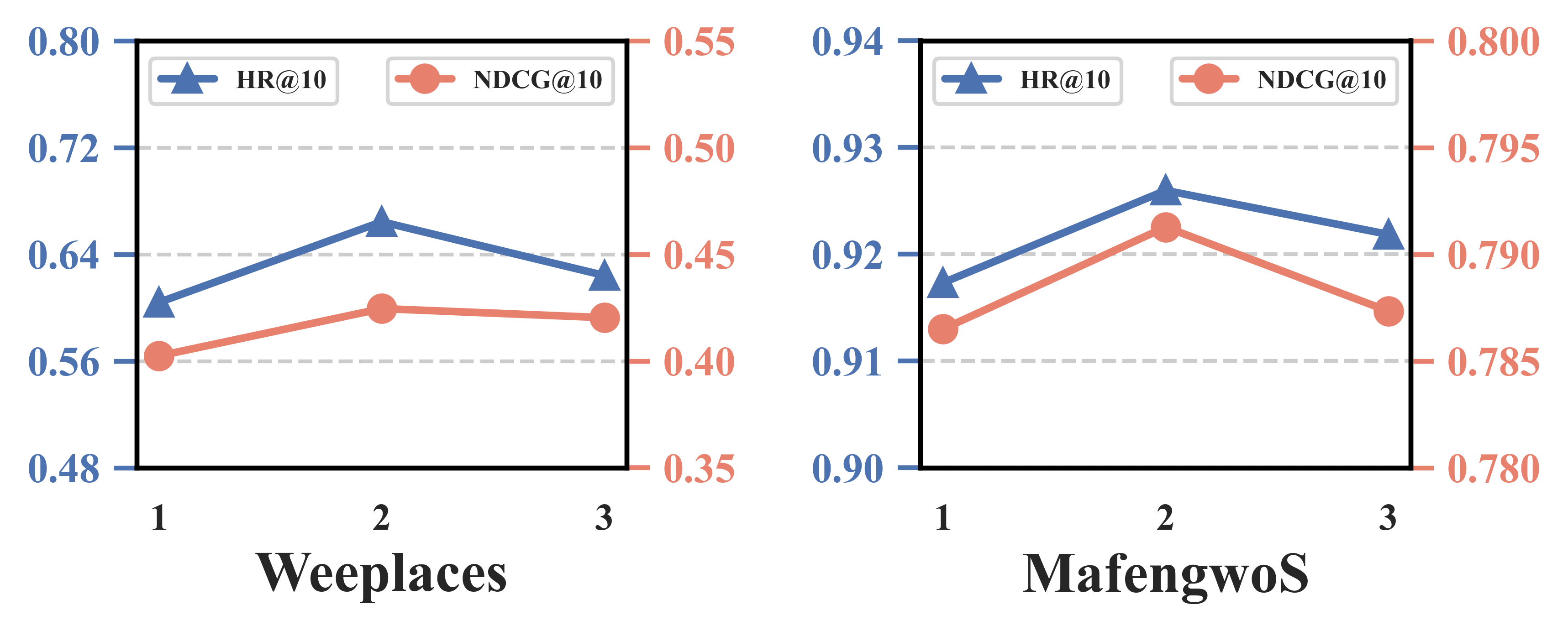} 
\caption{Impact of the order of meta-paths.}
\label{fig3}
\end{figure}

\subsubsection{Impact of LLM Types.}
We examine the impact of different base LLMs on AgentGR's performance using DeepSeek-V3, GLM-4-air, Qwen2.5, and GPT-4o. As shown in Figure 6, all variants perform well across all evaluation metrics, confirming the robustness of AgentGR's semantic meta-path guided CoP reasoning mechanism, semantic recognition modules of group topics and leadership, and multi-agent simulation strategy. Among them, DeepSeek-V3 and GPT-4o achieve the strongest and comparable performance, highlighting their advanced language understanding and reasoning capabilities. In addition, in scenarios where recommendation performance is the primary objective, AgentGR can be built on more powerful LLMs with higher computational cost to achieves stronger performance. In contrast, in cost-sensitive scenarios, AgentGR based on open-source LLMs remains practical, feasible, and effective.

\subsection{Case Study (RQ6)}
We conduct two case studies to further evaluate the effectiveness of AgentGR to capture semantic preference correlations and simulate dynamic group decision-making. In Case 1 (Figure~\ref{fig5}, left), the ground-truth destination, Saint Petersburg, is not connected to Group 25 within three-hop neighbors in the interaction graph, exposing the limitations of traditional methods. By incorporating semantic information, AgentGR identifies the group topic semantically aligned with Saint Petersburg and successfully recommends it, demonstrating its ability to capture semantic preferences beyond explicit interactions. In Case 2 (Figure~\ref{fig5}, right), Group 21 engages in a simulated multi-agent discussion. Member agents express item preferences based on their individual preferences, group history, topics, and leader cues. The recommendation agent aggregates the discussion outcome to generate a recommendation list, which is then reviewed and approved by a consensus judgment agent. All members converge on the group topic of Tibetan culture and natural landscapes, demonstrating AgentGR's capacity to simulate realistic consensus-building processes.

\subsection{Hyperparameter Analysis (RQ7)}
The performance of user preference learning is influenced by the order of meta-paths. We investigate the impact of the order $h$ within the range of [1, 2, 3], and the results are shown in Figure~\ref{fig3}. The results show that AgentGR benefits a larger $h$ across all datasets, achieving the best performance with $h=2$. This phenomenon underscores the capacity of AgentGR to capture high-order collaborative filtering signals and enhance recommendation performance. Moreover, an excessive order might introduce noise, which has a detrimental effect on recommendation performance.

\section{Conclusion}
In this paper, we propose a novel GR model, named AgentGR. It jointly captures collaborative-semantic user preferences for member-role-playing and simulates dynamic group interactions to reflect real-world group decision-making processes, thereby boosting recommendations. Specifically, we first introduce a semantic meta-path guided CoP reasoning mechanism that integrates high-order collaborative filtering signals and textual semantics to improve user preference profiles. We then recognize group topics and leadership to explicitly model the influencing factors in the group decision processes. Finally, we simulate group-level decision dynamics via two multi-agent simulation strategies for recommendations. Extensive experiments on two real-world datasets confirm the effectiveness and superiority of AgentGR.

\bibliographystyle{ACM-Reference-Format}
\bibliography{sample-base}










\end{document}